\documentstyle[twoside,psfig]{article}
\input{epsf.sty}

\catcode`\@=11
\long\def\@makefntext#1{
\protect\noindent \hbox to 3.2pt {\hskip-.9pt  
$^{{\eightrm\@thefnmark}}$\hfil}#1\hfill} 

\def\@makefnmark{\hbox to 0pt{$^{\@thefnmark}$\hss}} 

\def\ps@myheadings{\let\@mkboth\@gobbletwo
\def\@oddhead{\hbox{}
\rightmark\hfil\eightrm\thepage}   
\def\@oddfoot{}\def\@evenhead{\eightrm\thepage\hfil
\leftmark\hbox{}}\def\@evenfoot{}
\def\sectionmark##1{}\def\subsectionmark##1{}}

\oddsidemargin=\evensidemargin
\newcounter{sectionc}\newcounter{subsectionc}\newcounter{subsubsectionc}
\renewcommand{\section}[1] {\vspace{12pt}\addtocounter{sectionc}{1} 
\setcounter{subsectionc}{0}\setcounter{subsubsectionc}{0}\noindent 
   {\tenbf\thesectionc. #1}\par\vspace{5pt}}
\renewcommand{\subsection}[1] {\vspace{12pt}\addtocounter{subsectionc}{1} 
\setcounter{subsubsectionc}{0}\noindent 
   {\bf\thesectionc.\thesubsectionc. {\kern1pt \bfit #1}}\par\vspace{5pt}}
\renewcommand{\subsubsection}[1]{\vspace{12pt}\addtocounter{subsubsectionc}{1}
\noindent{\tenrm\thesectionc.\thesubsectionc.\thesubsubsectionc.
{\kern1pt \tenit #1}}\par\vspace{5pt}}
\newcommand{\nonumsection}[1] {\vspace{12pt}\noindent{\tenbf #1}
\par\vspace{5pt}}

\newcounter{appendixc}
\newcounter{subappendixc}[appendixc]
\newcounter{subsubappendixc}[subappendixc]
\renewcommand{\thesubappendixc}{\Alph{appendixc}.\arabic{subappendixc}}
\renewcommand{\thesubsubappendixc}
{\Alph{appendixc}.\arabic{subappendixc}.\arabic{subsubappendixc}}

\renewcommand{\appendix}[1] {\vspace{12pt}
        \refstepcounter{appendixc}
        \setcounter{figure}{0}
        \setcounter{table}{0}
        \setcounter{lemma}{0}
        \setcounter{theorem}{0}
        \setcounter{corollary}{0}
        \setcounter{definition}{0}
        \setcounter{equation}{0}
        \renewcommand{\thefigure}{\Alph{appendixc}.\arabic{figure}}
        \renewcommand{\thetable}{\Alph{appendixc}.\arabic{table}}
        \renewcommand{\theappendixc}{\Alph{appendixc}}
        \renewcommand{\thelemma}{\Alph{appendixc}.\arabic{lemma}}
        \renewcommand{\thetheorem}{\Alph{appendixc}.\arabic{theorem}}
        \renewcommand{\thedefinition}{\Alph{appendixc}.\arabic{definition}}
        \renewcommand{\thecorollary}{\Alph{appendixc}.\arabic{corollary}}
        \renewcommand{\theequation}{\Alph{appendixc}.\arabic{equation}}
        \noindent{\tenbf Appendix#1}\par\vspace{5pt}}
\newcommand{\subappendix}[1] {\vspace{12pt}
        \refstepcounter{subappendixc}
        \noindent{\bf Appendix \thesubappendixc. {\kern1pt \bfit #1}}
\par\vspace{5pt}}
\newcommand{\subsubappendix}[1] {\vspace{12pt}
        \refstepcounter{subsubappendixc}
        \noindent{\rm Appendix \thesubsubappendixc. {\kern1pt \tenit #1}}
\par\vspace{5pt}}

\newcommand{\textlineskip}{\baselineskip=13pt}
\newcommand{\smalllineskip}{\baselineskip=10pt}

\def\eightcirc{
\begin{picture}(0,0)
\put(4.4,1.8){\circle{6.5}}
\end{picture}}
\def\eightcopyright{\eightcirc\kern2.7pt\hbox{\eightrm c}} 

\newcommand{\copyrightheading}[1]
{\vspace*{-2.5cm}\smalllineskip{\flushleft
{\footnotesize Mathematical Models and Methods in 
                       Applied Sciences #1}\\
{\footnotesize $\eightcopyright$\, World Scientific Publishing
Company}\\
}}

\newcommand{\pub}[1]{{\begin{center}\footnotesize\smalllineskip 
#1\\ 
\end{center}
}}

\def\abstracts#1#2#3{{
\centering{\begin{minipage}{4.5in}\baselineskip=10pt\footnotesize
\parindent=0pt #1\par 
\parindent=15pt #2\par
\parindent=15pt #3
\end{minipage}}\par}} 

\renewenvironment{thebibliography}[1]
{\frenchspacing
\ninerm\baselineskip=11pt
\begin{list}{\arabic{enumi}.}
        {\usecounter{enumi}\setlength{\parsep}{0pt}     
\setlength{\leftmargin 12.7pt}{\rightmargin 0pt} 
         \setlength{\itemsep}{0pt} \settowidth
{\labelwidth}{#1.}\sloppy}}{\end{list}}

\newcounter{itemlistc}
\newcounter{romanlistc}
\newcounter{alphlistc}
\newcounter{arabiclistc}

\newcommand{\fcaption}[1]{
        \refstepcounter{figure}
        \setbox\@tempboxa = \hbox{\footnotesize Fig.~\thefigure. #1}
        \ifdim \wd\@tempboxa > 5in
           {\begin{center}
        \parbox{5in}{\footnotesize\smalllineskip Fig.~\thefigure. #1}
            \end{center}}
        \else
             {\begin{center}
             {\footnotesize Fig.~\thefigure. #1}
              \end{center}}
        \fi}

\newcommand{\tcaption}[1]{
        \refstepcounter{table}
        \setbox\@tempboxa = \hbox{\footnotesize Table~\thetable. #1}
        \ifdim \wd\@tempboxa > 5in
           {\begin{center}
        \parbox{5in}{\footnotesize\smalllineskip Table~\thetable. #1}
            \end{center}}
        \else
             {\begin{center}
             {\footnotesize Table~\thetable. #1}
              \end{center}}
        \fi}

\def\@citex[#1]#2{\if@filesw\immediate\write\@auxout
{\string\citation{#2}}\fi
\def\@citea{}\@cite{\@for\@citeb:=#2\do
{\@citea\def\@citea{,}\@ifundefined
{b@\@citeb}{{\bf ?}\@warning
{Citation `\@citeb' on page \thepage \space undefined}}
{\csname b@\@citeb\endcsname}}}{#1}}

\newif\if@cghi
\def\cite{\@cghitrue\@ifnextchar [{\@tempswatrue
\@citex}{\@tempswafalse\@citex[]}}
\def\citelow{\@cghifalse\@ifnextchar [{\@tempswatrue
\@citex}{\@tempswafalse\@citex[]}}
\def\@cite#1#2{{$\null^{#1}$\if@tempswa\typeout
{IJCGA warning: optional citation argument 
ignored: `#2'} \fi}}

\def\pmb#1{\setbox0=\hbox{#1}
\kern-.025em\copy0\kern-\wd0
\kern.05em\copy0\kern-\wd0
\kern-.025em\raise.0433em\box0}

\def\fnt#1#2{\footnotetext{\kern-.3em
{$^{\mbox{\scriptsize #1}}$}{#2}}}

\def\fpage#1{\begingroup
\voffset=.3in
\thispagestyle{empty}\begin{table}[b]\centerline{\footnotesize #1}
\end{table}\endgroup}

\def\runninghead#1#2{\pagestyle{myheadings}
\markboth{{\protect\footnotesize\it{\quad #1}}\hfill}
{\hfill{\protect\footnotesize\it{#2\quad}}}}
\font\tenrm=cmr10
\font\tenit=cmti10 
\font\tenbf=cmbx10
\font\bfit=cmbxti10 at 10pt
\font\ninerm=cmr9

\font\eightrm=cmr8

\def\qed{\hbox{${\vcenter{\vbox{ 
   \hrule height 0.4pt\hbox{\vrule width 0.4pt height 6pt
   \kern5pt\vrule width 0.4pt}\hrule height 0.4pt}}}$}}

\def\theequation{\thesectionc.\arabic{equation}}  

\begin{document}
\setlength{\textheight}{7.7truein}  

\runninghead
{Wigner-Poisson and nonlocal drift-diffusion model equations for semiconductor 
superlattices }
{L. L. Bonilla and R. Escobedo }

\normalsize\textlineskip
\thispagestyle{empty}
\setcounter{page}{1}

\copyrightheading{} {Vol.~15, No.~8 (2005) 1253--1272}

\vspace*{0.88truein}

\fpage{1}
\centerline{\bf WIGNER-POISSON AND NONLOCAL }
\baselineskip=13pt
\centerline{\bf DRIFT-DIFFUSION MODEL EQUATIONS}
\baselineskip=13pt
\centerline{\bf FOR SEMICONDUCTOR SUPERLATTICES}

\vspace*{0.37truein}
\centerline{\footnotesize L. L. BONILLA and R. ESCOBEDO
}
\vspace*{0.015truein}
\centerline{\footnotesize\it
Grupo de Modelizaci\'on y Simulaci\'on Num\'erica; Escuela Polit\'ecnica Superior, 
Universidad Carlos III de Madrid
}
\baselineskip=10pt
\centerline{\footnotesize \it Av de la Universidad 30,
28911 Legan\'{e}s, Madrid, Spain}
\baselineskip=10pt
\centerline{\footnotesize \it bonilla@ing.uc3m.es, escobedo@math.uc3m.es}
\vspace*{0.225truein}                                 
\pub{\today} 
\vspace*{0.21truein}
\abstracts{A Wigner-Poisson kinetic equation describing charge transport
in doped semiconductor superlattices is proposed. Electrons are 
supposed to occupy the lowest miniband, exchange of lateral momentum is 
ignored and the electron-electron interaction is treated in the Hartree approximation. 
There are elastic collisions with impurities and inelastic collisions with phonons, 
imperfections, etc. The latter are described by a modified BGK (Bhatnagar-Gross-Krook) 
collision model that allows for energy dissipation while yielding charge continuity. In
the hyperbolic limit, nonlocal drift-diffusion equations are derived systematically from the 
kinetic Wigner-Poisson-BGK system by means of the Chapman-Enskog method. The
nonlocality of the original quantum kinetic model equations implies that the derived 
drift-diffusion equations contain spatial averages over one or more superlattice periods.
Numerical solutions of the latter equations show self-sustained oscillations of the current
through a voltage biased superlattice, in agreement with known experiments.
}{}{}



\vspace*{1pt}\textlineskip 
\section{Introduction} 
\vspace*{-0.5pt}
\noindent
Nonlinear charge transport in semiconductor superlattices has been widely 
studied in the last decade \cite{Wac02,Bon02}. A superlattice (SL) is a 
convenient approximation to a quasi-one-dimensional crystal that was originally
proposed by Esaki and Tsu to observe Bloch oscillations, i.e., the periodic coherent 
motion of electrons in a miniband in the presence of an applied electric field. 
When the materials were grown, many interesting nonlinear phenomena were
observed, including self-oscillations of the current through the SL due to
motion of electric field pulses, multistability of stationary solutions, and so on 
\cite{Wac02,Bon02}. In the effort to describe quantitatively transport in these 
materials, a large number of quantum and semiclassical equations, hydrodynamic 
and drift-diffusion models have been proposed. However, systematic derivations of 
reduced balance equations (hydrodynamic or drift-diffusion equations) are scarce. 
Hydrodynamic equations have been derived by Lei and coworkers 
\cite{LTi84,Lei95} and their numerical solutions describe self-oscillations of the current 
in different regimes \cite{CLe99}. They write operator Heisenberg equations, and average 
them to obtain a hierarchy of equations for moments such as electron density, current and 
energy. These equations are then closed by using a Fermi-Dirac local equilibrium ansatz, 
which clearly does not amount to a systematic procedure, nor gives a clue to the validity of 
the resulting hydrodynamic equations. A similar closure procedure has been recently developed 
for the case of quantum particles in an arbitrary external three-dimensional potential by 
Degond and Ringhofer \cite{DRi03}, who have used the maximum entropy principle to 
select their local equilibrium. These authors employ an operator form of a local Maxwellian 
in their examples, which therefore do not obey the Pauli exclusion principle. Earlier, Huang 
and Wu \cite{HWu94} found a local Fermi-Dirac distribution function as a basis of their 
hydrodynamic equations by using a different definition of entropy. 

For SL drift-diffusion models, systematic derivation procedures do exist. Many 
authors have used the Einstein relation to write a field-dependent diffusion coefficient 
from the electron drift velocity. The latter was obtained by Ignatov and Shashkin 
\cite{ISh87} from a simple solution of a Bhatnagar-Gross-Krook (BGK) kinetic 
equation \cite{BGK54}. Recently, we have used the Chapman-Enskog method to derive 
a generalized drift-diffusion equation (GDDE) from a BGK-Poisson system \cite{BEP03}. 
Interestingly, the Einstein relation does not hold, except in a very particular high-temperature 
limit \cite{BEP03}. Earlier, Cercignani, Gamba and Levermore had used essentially the 
same method to obtain reduced balance equations for a semiclassical BGK-Poisson 
semiconductor system treating the band dispersion relation in the parabolic approximation 
\cite{CGL01}. The aim of the present paper is to extend these methods to quantum
kinetic equations. To this end, we shall derive a Wigner-Poisson system of equations for a 
SL having only one populated miniband, ignore changes in lateral momentum and model the 
collisions in the BGK approximation. The resulting quantum kinetic equation is
\begin{eqnarray} 
&&{\partial f\over \partial t} + \frac{i}{\hbar}\left[ {\cal E}_{1}\left(k + 
{1\over 2i}{\partial\over \partial x}\right) - {\cal E}_{1}\left(k - {1\over 2i}
{\partial\over \partial x}\right)\right]\, f \nonumber\\
&&\quad +  {ie\over \hbar} \,\left[W
\left(x+{1\over 2i}{\partial\over \partial k},t\right) - W\left(x-{1\over 
2i}{\partial\over \partial k},t\right) \right]\, f \nonumber\\
&&\quad = - \nu_{en}\,  \left(f - f^{FD}
\right) - \nu_{\rm imp}\, {f(x,k,t) - f(x,-k,t)\over 2}  ,  \label{1}\\
&&\varepsilon\, {\partial^2 W\over\partial x^2} = {e\over l}\, 
(n-N_{D}),  \label{2}\\   
&& n = { l\over 2\pi} \int_{-\pi/l}^{\pi/l} f(x,k,t) dk =
{ l\over 2\pi} \int_{-\pi/l}^{\pi/l} f^{FD}(k;n) dk,\quad \label{3}\\
&& f^{FD}(k;n) = {m^{*}k_{B}T\over\pi^2\hbar^2}\, \left\{
\tan^{-1}\left(\frac{\Gamma}{{\cal E}_{1}(k)}\right)\right.\nonumber\\
&& \quad\left. +\int_{0}^\infty \frac{\Gamma}{\left[E- {\cal E}_{1}(k)\right]^2+
\Gamma^2}\,\ln\left[1+ \exp\left({\mu - E\over k_{B}T}\right)\right]\,
dE\right\}. \label{4}
\end{eqnarray} 
Here $f$, $n$, $N_{D}$, ${\cal E}_{1}(k)$, $d_{B}$, $d_{W}$, $l=d_{B}+d_{W}$, $W$, 
$\varepsilon$, $m^*$, $k_{B}$, $T$, $\Gamma$, $\nu_{en}$, $\nu_{\rm imp}$ and $-e<0$ 
are the one-particle Wigner function, the 2D electron density, the 2D doping density, the 
miniband dispersion relation, the barrier width, the well width, the SL 
period, the electric potential, the SL permittivity, the effective mass of the electron, the 
Boltzmann constant, the lattice temperature, the energy broadening of the equilibrium 
distribution due to collisions, the frequency of the inelastic collisions responsible for energy
relaxation, the frequency of the elastic impurity collisions and the electron charge, 
respectively. The chemical potential $\mu$ is a function of $n$ resulting from solving 
equation (\ref{3}) with the integral of the collision-broadened 3D Fermi-Dirac 
distribution over the lateral components of the wave vector $(k,k_{y},k_{z})$, which is 
given by Eq.\ (\ref{4}). Notice that, following Ignatov and Shashkin \cite{ISh87}, we 
have not included the effects of the electric potential in our Fermi-Dirac distribution. As
discussed later, the model equations we use can be improved by including scattering processes 
with change of lateral momentum and an electric field-dependent local equilibrium. However
the resulting model could only be treated numerically and the qualitative features of our
derivation and nonlocal drift-diffusion equation would be lost in longer formulas.

The Wigner-Poisson-Bhatnagar-Gross-Krook (WPBGK) system (\ref{1}) to (\ref{4}) 
should be solved for a Wigner function which is $2\pi/l$-periodic in $k$ and satisfies
appropriate initial and boundary conditions. If we integrate (\ref{1}) over $k$ and use
the periodicity condition, only the first two terms in its left hand side yield nonzero 
contributions which, in fact provide the charge continuity equation as we shall see later. 
Equation (\ref{1}) contains two pseudo-differential operators, depends on the unknown
electric potential $W$ and is nonlocal in $x$. However, the fact that $f$ is $2\pi/l$-periodic 
in $k$ makes it possible to use a modification of the Chapman-Enskog method along the
lines of that used in References \cite{CGL01} and \cite{BEP03} for the semiclassical 
case and earlier in \cite{Bon00} and \cite{BSo01} for nonlinear Fokker-Planck 
equations. The result is a generalized drift-diffusion equation for the electric field which is 
nonlocal in space because it contains averages of the electric field over one or more SL 
periods. The nonlocality of the {\em quantum drift-diffusion equation} (QDDE) is a direct 
consequence of the nonlocality of the Wigner equation (\ref{1}). 

In this paper, we derive the WPBGK system, deduce the nonlocal QDDE by means of the
Chapman-Enskog method and solve it numerically to illustrate the resulting self-sustained
oscillations of the current through a finite superlattice subject to an appropriate dc voltage
bias. The rest of the paper is organized as follows. In Section 2 we derive the WPBGK 
system. In Section 3, we study the hyperbolic limit of the WPBGK system and use the 
Chapman-Enskog method to derive the reduced equation to leading order. This equation is 
hyperbolic, and we would like to regularize possible singularities. Diffusive terms are 
obtained in Section 3 by calculating the first order terms in the Chapman-Enskog expansion. 
The resulting QDDE is discussed in Section 4 and solved numerically for appropriate voltage 
bias conditions. Finally, Section 5 contains a discussion of our results.

\vspace*{1pt}\textlineskip 
\setcounter{equation}{0}
\section{Derivation of the quantum kinetic equation} 
\vspace*{-0.5pt}

\noindent
Let us consider an $n$-doped SL formed by periodic alternation of two different
semiconductors such as GaAs (quantum well W) and AlAs (barrier B). The SL cross section 
$S$ is much larger than $l^2$, the square of the SL period. For $n$-doped SLs, we can 
restrict ourselves to studying electronic transport in the conduction band of the SL. We 
shall assume that the SL is under a dc voltage bias, which is equivalent to an external 
electric field directed along the SL growth direction. The corresponding Hamiltonian is
\begin{eqnarray}
H = H_{0} + H_{e-e} + H_{sc}. \label{eq2a1}
\end{eqnarray}
We have separated the electron-electron interaction $H_{e-e}$ and other scattering processes
(impurity, phonon,...) $H_{sc}$ from the one-electron Hamiltonian $H_{0}$. Typically,
the electron-electron interaction is treated in the Hartree approximation. Then we can find
the spectrum of the Hamiltonian $H_{0} + H_{e-e}$ by solving a non-linear stationary
Schr\"odinger-Poisson system of equations. Their solutions yield a basis in which quantum
kinetic equations describing the scattering processes out of equilibrium can be written, as
shown below. The envelope wave function is \cite{Basbook88}
\begin{eqnarray}
\varphi_{\nu}({\bf x},{\bf k})\equiv \varphi_{\nu}(x,{\bf x}_{\perp},k,{\bf 
k}_{\perp}) = {1\over\sqrt{S}}\, e^{i {\bf k}_{\perp}\cdot {\bf x}_{\perp}} 
\varphi_{\nu}(x,k,{\bf k}_{\perp}).    \label{eq2a2}
\end{eqnarray}
At zero external field, $\varphi_{\nu}$ satisfies
\begin{eqnarray}
&& \left[ {\cal E}_{c} + V_{c}(x) -e W(x)- {\hbar^2\over 2}\, {\partial\over
\partial x} {1\over m(x)} {\partial\over\partial x} + {\hbar^2{\bf k}_{\perp}
^2\over 2\, m(x)}\right] \varphi_{\nu} = {\cal E}_{\nu}({\bf k})\, 
\varphi_{\nu} ,    \label{eq2a3}\\
&& \varepsilon(x)\, {\partial^2 W\over\partial x^2} = e\, 
[n_{0} - N_{3D}(x)].  \label{eq2a4}
\end{eqnarray}
Here ${\cal E}_{c}$ is the conduction band edge of material W (GaAs, well), $W(x)$ is 
the electric potential due to the electron-electron interaction, $N_{3D}(x)=N_{D}(x)/l$ is 
the 3D doping density, and
\begin{eqnarray}
m(x) = \left\{ \begin{array}{ll}
m_{W} & \mbox{if $x$ corresponds to a quantum well W}\\
m_{B} & \mbox{if $x$ corresponds to a barrier B}\end{array}\right.  
\label{eq2a5}\\
\varepsilon(x) = \left\{ \begin{array}{ll}
\varepsilon_{W} & \mbox{if $x$ corresponds to a quantum well W}\\
\varepsilon_{B} & \mbox{if $x$ corresponds to a barrier B}\end{array}\right. 
\label{eq2a6}
\end{eqnarray}
are the masses and permittivities of the well and barrier. If $V_{c}$ corresponds
to the conduction-band offset between the well and barrier material, we have 
\begin{eqnarray}
V_{c}(x) = \left\{ \begin{array}{ll}
0 & \mbox{if $x$ corresponds to a quantum well W}\\
V_{c} & \mbox{if $x$ corresponds to a barrier B.}\end{array}\right.   
\label{eq2a7}
\end{eqnarray}
Moreover, the 3D equilibrium electron density is 
\begin{eqnarray}
n_{0}(x)= {1\over Sl}\, \sum_{\nu,k,{\bf k}_{\perp}} |\varphi_{\nu}(x,k,{\bf 
k}_{\perp})|^2 n_{F}(\nu,k,{\bf k}_{\perp}),   \label{eq2a8}
\end{eqnarray} 
where $n_{F}$ denotes the Fermi function of the miniband $\nu$. The boundary conditions 
at the well-barrier interfaces are that $\varphi_{\nu}$ and  $m(x)^{-1}\partial
\varphi_{\nu}/\partial x$ are both continuous. The electronic spectrum is continuous, 
consisting of {\em minibands} of energies ${\cal E}_{\nu}(k;{\bf k}_{\perp})$ with 
$\nu = 1, 2, 3,...$ (doubly degenerate because of spin) with an associated basis of spatially 
extended Bloch wave functions $\varphi_{\nu}(x,k;{\bf k}_{\perp}) = e^{ikx} u_{\nu}
(x,k;{\bf k}_{\perp})$ \cite{Basbook88}. 

Although we can discuss the effects of scattering using the previous basis of 
electronic states that solve a non-linear Schr\"odinger-Poisson system, we will
for simplicity ignore the difference in mass and permittivity between barriers and 
wells and assume that the doping density is uniform. Of course in real devices, only the
central parts of the wells are doped and the effective masses are different in wells and
barriers \cite{Grabook95}. These differences can be treated with the formulation sketched 
here (see for example, the Galdrikian-Birnir work for GaAs quantum wells, AlGaAs
barriers and a nonuniform doping density \cite{GBi96}), but then we would have to rely 
more heavily in numerical solutions than we intend to do, while obtaning qualitatively similar 
results (the AlAs barriers are much larger than the AlGaAs barriers and therefore the
change in the energy levels due to electrostatic effects is much less important in the 
GaAs/AlAs SLs we consider in this paper than in Galdrikian and Birnir's work). With these 
simplifications, the model [equations (\ref{eq2a3})--(\ref{eq2a7})] becomes the 
Kronig-Penney model with 
$m_{W}=m_{B}= m^*$, $\varepsilon_{W}= \varepsilon_{B}= \varepsilon$, and 
$N_{3D}(x)=N_{3D}$ (constant). The Schr\"odinger-Poisson problem is simply 
$[H_{0}- {\cal E}_{\nu}({\bf k})]\,\varphi_{\nu}({\bf x},{\bf k}) =0$ with 
$n_{0}=N_{3D}$, and its solutions have the form
\begin{eqnarray}
\varphi_{\nu}({\bf x},{\bf k}) = {1\over\sqrt{S}} e^{i{\bf k}_{\perp}\cdot
{\bf x}_{\perp}} \varphi_{\nu}(x,k), \quad {\cal E}_{\nu}({\bf k})  = {\hbar^2
{\bf k}^2_{\perp}\over 2 m^*} + {\cal E}_{\nu}(k).   \label{eq2a9}
\end{eqnarray}
The Bloch functions are $2\pi/l$-periodic in $k$, satisfying the orthogonality condition
\begin{eqnarray}
\int_{-\infty}^\infty \varphi_{\mu}^*(x,k)\, \varphi_{\nu}(x,k')\,
dx =\delta_{\mu\nu}\delta(k- k'),  \label{eq2a10}
\end{eqnarray}
and the closure condition
\begin{eqnarray}
\int_{-\infty}^\infty \varphi_{\mu}^*(x,k)\, \varphi_{\nu}(x',k)\,
dk =\delta_{\mu\nu}\delta(x- x'),  \label{eq2a11}
\end{eqnarray}
provided the integral of $|\varphi_{\nu}|^2$ over one SL period is unity. 

Scattering different from electron-electron scattering is usually treated by writing equations
for the density matrix, its Wigner transform, or the non-equilibrium Green's function (NGF)
\cite{KBabook89,HJabook96,Wac02}. Whatever the chosen formulation, the equations 
for the one-electron functions depend on two-electron and higher functions, and we have the 
usual infinite hierarchy of coupled equations, which is well-known in classical kinetic 
theory. Typically, the hierarchy is closed by assuming some dependence of the two-electron 
functions on one-electron functions, which is suggested by perturbation theory in the limit
of weak scattering \cite{HJabook96}. Assuming weak scattering, the differences between 
the equations corresponding to the different formulations are small. The trouble is that the 
kinetic equations are often used in the opposite {\em hydrodynamic} limit, in which 
collisions due to scattering are dominant. Then the results of using different 
formalisms are not equivalent, which has resulted in some discussion and confusion.
In this paper, we shall not discuss the difference between formulations in a 
precise way. Instead, we shall write kinetic equations for the one-electron density matrix or
Wigner function leaving unspecified the collision terms as much as we can, and discuss
how to obtain reduced theories for electric field, electron density and current, and so on. 
These theories are easier to analyze and to solve numerically, and they are the ones commonly used to 
understand non-linear phenomena in SLs.

To find a kinetic equation, we start writing equations for the coefficients 
$a_{\nu,{\bf k}}(t)$ in the expansion of the wave function
\begin{eqnarray}
\psi({\bf x},t)= \sum_{\nu,{\bf k}} a_{\nu,{\bf k}}(t)\, \varphi_{\nu}({\bf x},{\bf 
k}) \equiv \sum_{\nu}\psi_{\nu}({\bf x},t).  \label{eq2a14}
\end{eqnarray}
If we ignore the scattering term $H_{sc}$ in equation (\ref{eq2a1}), the coefficients 
$a_{\nu,{\bf k}}(t)$ become
\begin{eqnarray}
i\hbar \frac{\partial }{\partial t}a_{\nu,{\bf k}} = {\cal E}_{\nu}({\bf k}) 
a_{\nu,{\bf k}} - e \sum_{\nu',{\bf k'}} \langle \nu {\bf k}|W|\nu' {\bf k'}\rangle
\, a_{\nu',{\bf k'}}.  \label{eq2a15}
\end{eqnarray}
The equations for the {\em band wave functions} $\psi_{\nu}$ of equation (\ref{eq2a14}) 
can be obtained from this equation after some algebra
\begin{eqnarray}
 i\hbar \frac{\partial }{\partial t}\psi_{\nu} &=& - \frac{\hbar^2}{2m^*}
\frac{\partial^2}{\partial {\bf x}_{\perp}^2}\psi_{\nu} + 
\sum_{m=-\infty}^\infty E_{\nu}(m)\,\psi_{\nu}(x+ml,{\bf x}_{\perp},t) 
\nonumber\\
&-& e \sum_{\nu'} \int \Phi_{\nu}({\bf x},{\bf x'})\, W({\bf x'}) \psi_{\nu'}({\bf 
x'},t)\, d{\bf x'},   \label{eq2a16}\\
\Phi_{\nu}({\bf x},{\bf x'}) &=& \sum_{{\bf k}} \varphi_{\nu}({\bf x},{\bf k})\,
\varphi_{\nu}^*({\bf x'},{\bf k}),    \label{eq2a17}\\
{\cal E}_{\nu}(k) &=& \sum_{m=-\infty}^\infty E_{\nu}(m)\, e^{imkl}.   
\label{eq2a18}
\end{eqnarray}
Notice that equation (\ref{eq2a2}) implies
\begin{eqnarray}
\Phi_{\nu}({\bf x},{\bf x'})= \delta({\bf x}_{\perp}-{\bf x'}_{\perp})\,
\phi_{\nu}(x,x'), \nonumber\\
\phi_{\nu}(x,x')= \sum_{k} \varphi_{\nu}(x,k)\,\varphi_{\nu}^*(x',k), 
\label{eq2a19}
\end{eqnarray}
and the closure condition in equation (\ref{eq2a11}) yields
\begin{eqnarray}
\sum_{\nu}\Phi_{\nu}({\bf x},{\bf x'})= \delta({\bf x}-{\bf x'}).  \label{eq2a20}
\end{eqnarray}
Thus $\Phi_{\nu}({\bf x},{\bf x'})$ can be considered as the projection of the delta
function $\delta({\bf x}-{\bf x'})$ onto the band $\nu$.

After second quantization, the band density matrix is defined by
\begin{eqnarray}
\rho_{\mu,\nu}({\bf x},{\bf y},t) = \langle \psi^{\dag}_{\mu}({\bf x},t)\, 
\psi_{\nu} ({\bf y},t)\rangle,     \label{eq2a21}
\end{eqnarray}
so that the 2D electron density is (the factor 2 is due to spin degeneracy)
\begin{eqnarray}
n({\bf x},t) = 2l\,\sum_{\mu,\nu}\langle \psi^{\dag}_{\mu}({\bf x},t)\,\psi_{\nu} 
({\bf x},t)\rangle = 2l\,\sum_{\mu,\nu}\rho_{\mu,\nu}({\bf x},{\bf x},t).  
\label{eq2a22}
\end{eqnarray}
Using equations (\ref{eq2a21}) and (\ref{eq2a22}), we can derive the following evolution 
equation for the band density matrix 
\begin{eqnarray}
&& i\hbar \frac{\partial }{\partial t}\rho_{\mu,\nu} + \frac{\hbar^2}{2m^*}
\left(\frac{\partial^2}{\partial {\bf y}_{\perp}^2}- \frac{\partial^2}{
\partial {\bf x}_{\perp}^2}\right)\rho_{\mu,\nu} \nonumber\\
&& - \sum_{m=-\infty}^\infty [E_{\nu}(m) 
\rho_{\mu,\nu}({\bf x},y+ml,{\bf y}_{\perp},t) - E_{\mu}(m) \rho_{\mu,\nu}(x-ml,
{\bf x}_{\perp},{\bf y},t)] \nonumber\\
&&+ e \sum_{\nu'} \int W({\bf z})\, [\Phi_{\nu}({\bf y},
{\bf z}) \rho_{\mu\nu'}({\bf x},{\bf z},t) - \Phi_{\mu}({\bf z},{\bf x})
\rho_{\nu'\nu}({\bf z},{\bf y},t)] = Q[\rho],     \label{eq2a23}
\end{eqnarray}
with $Q[\rho]\equiv 0$ in the absence of scattering. The Hartree potential satisfies the 
Poisson equation
\begin{eqnarray}
\varepsilon\, {\partial^2 W\over\partial x^2} = {e\over l}\, (n - N_{D}) 
\label{eq2a24}
\end{eqnarray}
(recall that $e>0$ and that $W$ is the electric potential. The charge density is thus equal to 
minus the right hand side of this equation). When considering scattering, the right hand side 
of equation (\ref{eq2a23}) is equal to a
non-zero functional of the band density matrix $Q[\rho]$, whose form depends on the 
closure assumption we have made to close the density matrix hierarchy. In the 
semiclassical limit, the kernel of the collision term $Q[\rho]$ is usually found by using
perturbation theory in the impurity potential, electron-phonon interaction, etc. For the time 
being, we shall not try to formulate collision models. Instead and
in order to make contact with the kinetic equations in the semiclassical limit, we shall rewrite 
equation (\ref{eq2a23}) in terms of the band Wigner function 
\begin{eqnarray}
w_{\mu,\nu}({\bf x},{\bf k},t)= \int \rho_{\mu,\nu}\left({\bf x}+{1\over 2}{\bf 
\xi},{\bf x}-{1\over 2}{\bf \xi},t\right)\, e^{i{\bf k}\cdot {\bf \xi}} d{\bf\xi}. 
\label{eq2a25}
\end{eqnarray}
The evolution equation for the Wigner function is
\begin{eqnarray}
&&\frac{\partial }{\partial t}w_{\mu,\nu} + \frac{\hbar {\bf k}_{\perp}}{m^*}
\cdot \frac{\partial}{\partial {\bf x}_{\perp}}w_{\mu,\nu} 
+\frac{i}{\hbar} \sum_{m=-\infty}^\infty e^{imkl} \left[E_{\nu}(m) 
w_{\mu,\nu}\left(x+{ml\over 2},{\bf x}_{\perp},{\bf k},t\right)\right.
\nonumber\\
&& \left.- E_{\mu}(m) w_{\mu,\nu}\left(x-{ml\over 2},{\bf x}_{\perp},{\bf k},t
\right)\right] 
+ \frac{ie}{\hbar} \sum_{\nu'} \int \left[W\left(z+\frac{1}{2i}
\frac{\partial}{\partial k},{\bf x}_{\perp}\right) \right.\nonumber\\
&& \times \phi_{\mu}(z,x)\, e^{ik(x-z)} 
w_{\nu',\nu}\left({x+z\over 2},{\bf x}_{\perp},{\bf k},t\right) 
- W\left(z-\frac{1}{2i}\frac{\partial}{\partial k},{\bf x}_{\perp}\right)
\nonumber\\
&&\left.\times \phi_{\nu}(x,z)\, e^{-ik(x-z)}
w_{\mu,\nu'}\left({x+ z\over 2},{\bf x}_{\perp},{\bf k},t\right)\right]\, dz = 
Q_{\mu,\nu}[w],        \label{eq2a27}
\end{eqnarray}
in which the collision term is again left unspecified. Notice that the 2D electron density is
\begin{eqnarray}
n({\bf x},t) = {2l\over 8\pi^3}\sum_{\mu,\nu}\int w_{\mu,\nu}({\bf x},{\bf k},t)\, 
d{\bf k},     \label{eq2a28}
\end{eqnarray}
because of equation (\ref{eq2a22}) and the definition in equation (\ref{eq2a25}). From 
equations (\ref{eq2a27}) and (\ref{eq2a28}), we obtain the charge continuity equation
\begin{eqnarray}
{e\over l}\, \frac{\partial n}{\partial t} + \frac{\partial}{\partial {\bf x}}
\cdot {\bf J} = 0,  \label{eq2a29}\\
{\bf J}_{\perp} = {2e\over 8\pi^3}\int \frac{\hbar {\bf k}_{\perp}}{m^*}\, 
\sum_{\mu,\nu} w_{\mu,\nu}({\bf x},{\bf k},t)\, d{\bf k}, \label{eq2a30}\\
\frac{\partial J}{\partial x} = \frac{ie}{4\pi^3\hbar}
\sum_{\mu,\nu,m} \int e^{imkl} \left[E_{\nu}(m) 
w_{\mu,\nu}\left(x+{ml\over 2},{\bf x}_{\perp},{\bf k},t\right)\right.
\nonumber\\
\quad\quad\left.- E_{\mu}(m) w_{\mu,\nu}\left(x-{ml\over 2},{\bf x}_{\perp},
{\bf k},t\right)\right]\, d{\bf k},    \label{eq2a31}
\end{eqnarray}
provided our collision model satisfies $\int\sum_{\mu,\nu}Q_{\mu,\nu} d{\bf k}=0$.

A related formulation of the band Wigner functions (without collision terms) is due to 
Demeio {\em et al} \cite{DBBpb02}. One difficulty with our formulation is that the Wigner 
function in equation (\ref{eq2a25}) is not $2\pi/l$-periodic in $k$. This can be 
corrected by using the following definition 
\begin{eqnarray}
f_{\mu,\nu}({\bf x},{\bf k},t) \equiv \sum_{s=-\infty}^\infty
w_{\mu,\nu}\left({\bf x},k+\frac{2\pi s}{l},{\bf k}_{\perp},t\right) 
= \sum_{j=-\infty}^\infty e^{ijkl}   l    \nonumber\\
\quad\times \int \rho_{\mu,\nu}\left(x+{jl\over 2},{\bf 
x}_{\perp}+{1\over 2}{\bf \xi}_{\perp},x-{jl\over 2},{\bf x}_{\perp}-{1\over 
2}{\bf \xi}_{\perp},t\right)\, e^{i{\bf k}_{\perp}\cdot {\bf \xi}_{\perp}} 
d{\bf\xi}_{\perp}.   \label{eq2a32}
\end{eqnarray}
To derive this equation, we have used the identity 
\begin{eqnarray}
\sum_{j=-\infty}^\infty \delta(\xi-jl) = {1\over l}\, \sum_{s=-\infty}^\infty
e^{i2\pi\xi s/l} ,     \label{eq2a33}
\end{eqnarray}
together with the definition of equation (\ref{eq2a25}). From equations (\ref{eq2a28}) and 
(\ref{eq2a32}), we obtain the 2D electron density in terms of $f_{\mu,\nu}$
\begin{eqnarray}
n({\bf x},t) = {2l\over 8\pi^3}\sum_{\mu,\nu}\int_{-\pi/l}^{\pi/l}\int 
f_{\mu,\nu}({\bf x},{\bf k},t)\, dk\, d{\bf k}_{\perp}.     \label{eq2a34}
\end{eqnarray}
Similarly, the transversal current density can be obtained from equations (\ref{eq2a30}) and 
(\ref{eq2a32})
\begin{eqnarray}
{\bf J}_{\perp} = {2 e\over 8\pi^3}\int_{-\pi/l}^{\pi/l}\int \frac{\hbar 
{\bf k}_{\perp}}{m^*}\, \sum_{\mu,\nu} f_{\mu,\nu}({\bf x},{\bf k},t)\, dk\,
d{\bf k}_{\perp}.   \label{eq2a35}
\end{eqnarray}
The current density along the growth direction has the form
\begin{eqnarray}
J = {2 e\hbar\over 8\pi^3m^*}\sum_{\mu,\nu,s}\int_{-\pi/l}^{\pi/l}\int 
\left(k+\frac{ 2\pi s}{l}\right) w_{\mu,\nu}\left({\bf x},k+\frac{ 2\pi s}{l}
,{\bf k}_{\perp},t\right) dk\, d{\bf k}_{\perp},  \label{eq2a36}
\end{eqnarray}
from which we can also derive equation (\ref{eq2a31}).

The definition of periodic band Wigner function is related to that adopted by Bechouche 
{\em et al} \cite{BMPcpam01}. These authors have rigorously proved that the collisionless 
Wigner-Poisson equations for a crystal become the crystal Vlasov-Poisson equations in the 
semiclassical limit assuming that the initial conditions are concentrated in isolated bands. 
Scattering other than electron-electron scattering is not considered in these works. A
similar work for a stratified material can be found in \cite{BPomm00}.

To find a Wigner-Poisson description of transport in a single miniband, we sum all the 
Wigner equations (\ref{eq2a27}) over the band indices and use the closure condition in 
equation (\ref{eq2a20}), so as to find an equation for $w({\bf x},{\bf k},t)= 
\sum_{\mu,\nu}w_{\mu,\nu}({\bf x},{\bf k},t)$
\begin{eqnarray}
&&\frac{\partial }{\partial t}w + \frac{\hbar {\bf k}_{\perp}}{m^*}
\cdot \frac{\partial}{\partial {\bf x}_{\perp}}w
+\frac{i}{\hbar} \sum_{m=-\infty}^\infty e^{imkl} \sum_{\mu,\nu}
\left[E_{\nu}(m) w_{\mu,\nu}\left(x+{ml\over 2},{\bf x}_{\perp},{\bf k},t
\right)\right.
\nonumber\\
&& \left.- E_{\mu}(m) w_{\mu,\nu}\left(x-{ml\over 2},{\bf x}_{\perp},{\bf k},t
\right)\right] + \frac{ie}{\hbar} \left[W\left(x+\frac{1}{2i}
\frac{\partial}{\partial k},{\bf x}_{\perp}\right) \right.\nonumber\\
&&\left. - W\left(x-\frac{1}{2i}
\frac{\partial}{\partial k},{\bf x}_{\perp}\right)\right]\, w = 
\sum_{\mu,\nu}Q_{\mu,\nu}[w].        \label{eq2b1}
\end{eqnarray}
Let us now assume that only the first miniband is populated and that there are no transitions
between minibands, $w({\bf x},{\bf k},t)\approx w_{1,1}({\bf x},{\bf k},t)$. 
This approximation is commonly used when describing strongly coupled SLs with wide 
minibands. Then equation (\ref{eq2b1}) becomes
\begin{eqnarray}
&&\frac{\partial }{\partial t}w + \frac{\hbar {\bf k}_{\perp}}{m^*}
\cdot \frac{\partial}{\partial {\bf x}_{\perp}}w
+\frac{i}{\hbar} \sum_{m=-\infty}^\infty e^{imkl} 
E_{1}(m)\, \left[w\left(x+{ml\over 2},{\bf x}_{\perp},{\bf k},t\right)\right.
\nonumber\\
&& \left.- w\left(x-{ml\over 2},{\bf x}_{\perp},{\bf k},t\right)\right] 
+ \frac{ie}{\hbar} \left[W\left(x+\frac{1}{2i}
\frac{\partial}{\partial k},{\bf x}_{\perp}\right) \right.\nonumber\\
&&\left. - W\left(x-\frac{1}{2i}
\frac{\partial}{\partial k},{\bf x}_{\perp}\right)\right]\, w = 
Q_{1,1}[w].        \label{eq2b2}
\end{eqnarray}
This yields the following equation for the periodic Wigner function in 
equation (\ref{eq2a32})
\begin{eqnarray}
&&\frac{\partial }{\partial t}f + \frac{\hbar {\bf k}_{\perp}}{m^*}
\cdot \frac{\partial}{\partial {\bf x}_{\perp}}f
+\frac{i}{\hbar} \sum_{m=-\infty}^\infty e^{imkl} 
E_{1}(m)\, \left[f\left(x+{ml\over 2},{\bf x}_{\perp},{\bf k},t\right)\right.
\nonumber\\
&& \left.- f\left(x-{ml\over 2},{\bf x}_{\perp},{\bf k},t\right)\right] 
+ \frac{ie}{\hbar} \left[W\left(x+\frac{1}{2i}
\frac{\partial}{\partial k},{\bf x}_{\perp}\right) \right.\nonumber\\
&&\left. - W\left(x-\frac{1}{2i}
\frac{\partial}{\partial k},{\bf x}_{\perp}\right)\right]\, f = 
Q[f].        \label{eq2b3}
\end{eqnarray}
The dispersion relation ${\cal E}_{1}(k)$ is an even periodic function of $k$ with period 
$2\pi/l $ that can be written as ${\cal E}_{1}(k) = \Delta_1 [1-\cos (kl)]/2$ plus a 
constant in the tight-binding approximation ($\Delta_1$ denotes the width of the first miniband). 
Moreover, the field ${\bf F}=\partial W/\partial {\bf x}$ (note that the real 
electric field is $-{\bf F}$) satisfies
\begin{eqnarray}
&& \varepsilon\,\left( {\partial F\over\partial x}  + {\partial\over\partial 
{\bf x}_{\perp}}\cdot {\bf F}_{\perp} \right) = {e\over l}\, (n-N_{D}),  
\label{eq2b4}\\   
&& n(x,{\bf x}_{\perp},t) = { l\over 4\pi^3} \int_{-\pi/l}^{\pi/l} 
\int_{-\infty}^\infty \int_{-\infty}^\infty f(x,{\bf x}_{\perp},k,{\bf 
k}_{\perp},t) dk\, d{\bf k}_{\perp}.    \label{eq2b5}
\end{eqnarray} 

We want to explicitly derive reduced balance equations from the kinetic equation. 
For this purpose, we need a sufficiently simplified description of scattering. 
Scattering processes such as phonon scattering change the energy and momentum of the 
electrons leading the distribution function toward thermal equilibrium. We can describe
these processes by a BGK collision model \cite{BGK54} 
\begin{eqnarray} 
&&Q_{en}[f]= - \nu_{en}\, (f - f^{FD}),     \label{eq2b6}\\
&& f^{FD}({\bf k};n) = \int_{0}^\infty \frac{\Gamma/\pi}{\left[E-
{\cal E}_{1}(k) - {\hbar^2{\bf k}_{\perp}^2\over 2m^*}\right]^2+\Gamma^2}\,
\frac{dE}{ 1 + \exp\left({ E- \mu\over k_{B}T}\right)} ,  \label{eq2b7}\\
&& n({\bf x},t) = { l\over 4\pi^3} \int_{-\pi/l}^{\pi/l} \int_{-
\infty}^{\infty}\int_{-\infty}^{\infty} f^{FD}(k,{\bf k}_{\perp};n) dk d{\bf 
k}_{\perp}. \label{eq2b8}
\end{eqnarray} 
Here $\Gamma$ measures the finite width of the spectral function in thermal equilibrium due 
to scattering \cite{Wac02}. As $\Gamma\to 0$, the first factor in equation (\ref{eq2b7})
becomes a delta function, and we recover the usual Fermi-Dirac distribution function with
a chemical potential $\mu$. The chemical potential $\mu=\mu(x,{\bf x}_{\perp},t)$ is a 
function of the exact electron density $n$ of equation~(\ref{eq2b5}) that is calculated by 
solving equation~(\ref{eq2b8}). With these definitions, the integral of $Q_{en}[f]$ over 
momentum vanishes, and the equation of charge continuity holds, as we shall show below. 
Notice that we have not included the electric potential in (\ref{eq2b7}). Then the 
equilibrium Wigner function does not include the $\hbar^2$ corrections to the semiclassical 
Fermi-Dirac distribution corresponding to the effective Hamiltonian ${\cal E}_{1}(-i
\partial/\partial x) - (\hbar^2/2m^*) \partial^2/\partial {\bf x}^2_{\perp}- e
W(x)$ (recall that the electron charge is $-e$): these corrections vanish if we set $W=0$ 
\cite{DRi03,Gar94}. Omitting the electric potential $W$ is certainly an imperfection of 
our BGK collision model, which could perhaps be corrected using field-dependent collision 
models as in \cite{DRi03}. However, the resulting technical complications would encumber 
our derivation of drift-diffusion type equations. Thus we prefer to adopt the same line as in 
previous works \cite{ISh87,BEP03} and leave the study of field-dependent and broadened 
collision models for the future.

Other processes, such as impurity scattering, conserve the energy of the electron, change only 
its momentum, and also preserve charge continuity. Gerhardts \cite{Gerprb93} used the 
following model 
\begin{eqnarray} 
Q_{\rm imp}[f] &=& - {\tilde{\nu}_{\rm imp}\over 4\pi^3}\,\int_{-\pi/l}^{\pi/l} 
\int_{-\infty}^{\infty}\int_{-\infty}^{\infty} \delta[E(k,{\bf k}_{\perp})- 
E(k',{\bf k}'_{\perp})] \nonumber\\
&\times & [f(k,{\bf k}_{\perp}) - f(k',{\bf k}'_{\perp})]\, dk'\, 
d{\bf k}'_{\perp},        \label{eq2b9}
\end{eqnarray} 
which can be rewritten as a relaxation to a weighted, energy-conserving mean value of the
Wigner function: $Q_{\rm imp}[f]= - \nu_{G}\, (f - \Phi_{f}[E({\bf k})])$, 
provided $\Phi_{f}[E]=\int \delta[E-E({\bf k}')] f({\bf k}') d^3{\bf k}'/[4\pi^3 
g(E)]$, $g(E)=\int \delta[E-E({\bf k}')]\, d^3{\bf k}'/(4\pi^3)$ is the
density of states, and $\nu_{G}=\tilde{\nu}_{\rm imp} g(E({\bf k}))$. Note that we 
have dropped the dependence of the Wigner function on space and time. This collision term 
couples the vertical motion of the electron to the lateral degrees of freedom. For SLs, 
scattering processes involving acoustic phonons, impurities and interface roughness may 
modify the lateral momentum. Liu {\em et al} \cite{LBA91} have calculated and measured 
the scattering times in weakly coupled double barrier heterostructures and obtained
energies smaller than 10 meV. This energy is also a generous upper bound for the energy 
exchanged in impurity scattering processes in strongly coupled SLs, as indicated in Wacker's 
review (his smallest scattering time is 0.0666 ps\cite{Wac02}). In any case, 10 meV is 
much less than the typical energy exchange in the direction parallel to the electric field, of 
more than 1 eV (which can be estimated from the width of the Brillouin zone, $2\pi/l$). 
Thus, we may reasonably assume that the variation of the energy in the lateral direction is 
negligible, $E(k,{\bf k}_{\perp})-E(k',{\bf k}'_{\perp})\approx {\cal E}_{1}(k)- 
{\cal E}_{1}(k')$, and therefore
\begin{eqnarray} 
&& Q_{\rm imp}[f]\approx - {\tilde{\nu}_{\rm imp}\over 2\pi}\,\int_{-\pi/l}^{\pi/l}
\delta[{\cal E}_{1}(k)- {\cal E}_{1}(k')]\, [f(k) - f(k')]\, dk' \nonumber\\
&& = -{\tilde{\nu}_{\rm imp}\over 2\pi |{\cal E}'_{1}(k)|}\, [f(k)- f(-k)]\equiv 
- {\nu_{\rm imp}\over 2}\, [f(k)- f(-k)],       \label{eq2b10}\\
&& f(x,k,t) = {1\over 2\pi^2 S}\int\int f(x,{\bf x}_{\perp},k,{\bf k'}_{\perp},t)\, 
d{\bf k'}_{\perp}d{\bf x}_{\perp}.   \label{eq2b11}
\end{eqnarray} 
Here ${\cal E}'_{1}(k)= d{\cal E}_{1}/dk$. The approximate collision term in 
equation~(\ref{eq2b10}) was introduced by Ktitorov {\em et al} \cite{KSSspss71}. In 
terms of the 1D Wigner function in equation (\ref{eq2b11}), we can derive from equations 
(\ref{eq2b3})--(\ref{eq2b11}) the 1D WPBGK system (\ref{1})-(\ref{4}) (which is 
independent of ${\bf x}_{\perp}$ if we assume that the initial Wigner function does not 
depend on ${\bf x}_{\perp}$). The Gerhardts collision model is still better than the 
approximation (\ref{eq2b10}) because his collision model explains the observed temperature 
dependence of the field at which the drift velocity has its lowest local maximum 
\cite{Gerprb93}. However the simpler 1D WPBGK model is much more convenient for 
analytical calculations. The semiclassical limit of the WPBGK system has been analyzed in 
\cite{BEP03} in order to derive a reduced drift-diffusion model in the hydrodynamic limit. 
Particular solutions of the semiclassical model have been found by different authors. Ignatov 
and Shashkin \cite{ISh87} found the stationary, space-independent semiclassical solution of 
equation~(\ref{1}) for a Boltzmann distribution function, the {\em Boltzmann limit} 
of equation~(\ref{4}), and studied its linear stability to plane wave disturbances in 
the field.

Before we proceed, it is convenient to derive the charge continuity equation and a form of
Amp\`ere's law for the current density from equation (\ref{1}). Since the Wigner 
function is periodic in $k$, we can write the second and third terms on the right hand side of 
equation (\ref{1}) in terms of its Fourier series
\begin{equation}
f(x,k,t) = \sum_{j=-\infty}^{\infty} f_{j}(x,t)\, e^{ijkl}.\label{eq2b16}
\end{equation}
We have
\begin{eqnarray} 
W\left(x\pm {1\over 2i}{\partial\over \partial k},t\right) f= \sum_{j=
-\infty}^{\infty} W\left(x\pm {jl\over 2},t\right)\, f_{j}\, e^{ijkl},
\label{eq2b17}
\end{eqnarray} 
and therefore,
\begin{eqnarray} 
W\left(x+ {1\over 2i}{\partial\over \partial k},t\right) f -
W\left(x- {1\over 2i}{\partial\over \partial k},t\right) f = \nonumber\\
\sum_{j= -\infty}^{\infty} [W(x+jl/2,t) - W(x-jl/2,t)]\, f_{j} e^{ijkl}
\nonumber\\
=\sum_{j= -\infty}^{\infty} jl \langle F\rangle_{j}\, f_{j}\, e^{ijkl}. 
\label{eq2b18}
\end{eqnarray} 
Here we have defined the average
\begin{eqnarray} 
\langle F\rangle_{j}(x,t) = {1\over jl} \int_{-jl/2}^{jl/2} F(x+s,t)\, ds.
\label{eq2b19}
\end{eqnarray} 
Note that differentiating an average, we obtain a finite difference 
\begin{eqnarray} 
{\partial\over \partial x}\langle g\rangle_{j}=\langle {\partial g\over 
\partial x}\rangle_{j} = {g(x+jl/2) - g(x-jl/2)\over jl} .   \label{eq2b20}
\end{eqnarray} 
Then the second term in equation (\ref{1}) is 
\begin{eqnarray} 
&&{i\over\hbar} \sum_{j= -\infty}^{\infty} \left[f\left(x+{jl\over 2},k,t\right) 
- f\left(x-{jl\over 2},k,t\right)\right] E_{1}(j)
e^{ijkl}\nonumber\\
&& = \sum_{j= -\infty}^{\infty} {ijl\over\hbar}\, e^{ijkl} E_{1}(j)\,
{\partial\over \partial x}\langle f\rangle_{j},  \label{eq2b21}
\end{eqnarray} 
which in the case of the tight-binding dispersion relation ${\cal E}_{1}(k)=\Delta_1\, (1-
\cos kl)/2$ becomes $v(k)\,\partial
\langle f\rangle_{1}/ \partial x$, with the usual miniband group velocity 
\begin{eqnarray} 
v(k)\equiv {1\over\hbar}\, \frac{d{\cal E}_{1}}{dk}= {\Delta_1 l\over 2\hbar}
\sin(kl).  \label{eq2b22}
\end{eqnarray} 
Inserting equations (\ref{eq2b18}) and (\ref{eq2b21}) into equation (\ref{1}), we obtain 
the following equivalent form of the Wigner equation, which is particularly suitable for 
treating SL problems
\begin{eqnarray} 
{\partial f\over \partial t} + \sum_{j= -\infty}^{\infty} {i jl\over\hbar}\, 
e^{ijkl} \left( E_{1}(j)\, {\partial\over \partial x}\langle f\rangle_{j}
+ e\, \langle F\rangle_{j}\, f_{j} \right) \nonumber\\
= - \nu_{en}\,  \left(f - f^{FD} \right) - \nu_{\rm imp}\, {f(x,k,t) - f(x,-k,t)\over 2}  .  
\label{eq2b23}
\end{eqnarray} 
We now integrate this equation over $k$, thereby getting the charge continuity equation
\begin{eqnarray} 
{\partial n\over \partial t} + {\partial\over \partial x}\, \sum_{j= 1}^{\infty
}{2jl\over\hbar} \left\langle \mbox{Im}(E_{1}(-j) f_{j})\right\rangle_{j}= 0.  
\label{eq2b24}
\end{eqnarray} 
We can eliminate the electron density from equation (\ref{eq2b24}) by using the Poisson 
equation (\ref{2}) and integrating the result over $x$, thereby obtaining the non-local 
Amp\`ere's law
\begin{eqnarray} 
\varepsilon\, {\partial F\over \partial t} + {2e\over\hbar}\,
\sum_{j= 1}^{\infty} j\langle \mbox{Im} (E_{1}(-j) f_{j})\rangle_{j} 
= J(t).   \label{eq2b25}
\end{eqnarray} 
Here $J(t)$ is the total current density. Equations (\ref{eq2b23}), (\ref{eq2b24}) and 
(\ref{eq2b25}) are spatially non-local versions of the corresponding semiclassical equations. 
The charge continuity and Amp\`ere's equations have their traditional form as derived from 
semiclassical Boltzmann equations, except that the electron current is averaged over the 
SL periods. This non-locality will be transmitted to the QDDE.

\vspace*{1pt}\textlineskip 
\setcounter{equation}{0}
\section{Derivation of the QDDE in the hyperbolic limit of the kinetic equation} 
\vspace*{-0.5pt}

\noindent 
To derive the QDDE, we shall assume that the electric field contribution in Eq.\ 
(\ref{eq2b23}) is comparable to the collision terms and that they dominate the other terms 
({\em the hyperbolic limit}) \cite{BEP03}. Let $v_{M}$ and $F_{M}$ be electron velocity 
and field scales typical of the macroscopic phenomena described by the sought balance 
equation; for example, let them be the positive values at which the (zeroth order) drift 
velocity reaches its maximum. In the hyperbolic limit, the time $t_{0}$ it takes an electron 
with speed $v_{M}$ to traverse a distance $x_{0}=\varepsilon F_{M}l/(eN_{D})$, over 
which the field variation is of order $F_{M}$, is much longer than the mean free time 
between collisions, $\nu_{en}^{-1}\sim \hbar/(eF_{M}l)=t_{1}$. We therefore define the
{\em small parameter} $\lambda=t_{1}/t_{0}=\hbar v_{M}N_{D}/(\varepsilon 
F_{M}^2 l^2)$ and formally multiply the first two terms on the left side of (\ref{1}) 
and of (\ref{eq2b23}) by $\lambda$ \cite{BEP03}. After obtaining the number of
desired terms, we set $\lambda =1$. 

The solution of equation~(\ref{eq2b23}) for $\lambda =0$ is the stationary 
space-independent solution that is easily found as a Fourier series
\begin{eqnarray} 
f^{(0)}(k;F) = \sum_{j=-\infty}^{\infty} f^{(0)}_{j} e^{ijkl}, \quad
f^{(0)}_{j} = {1-ij {\cal F}_{j}/\tau_{e}\over 1 + j^2 {\cal F}_{j}^{2}}\, f^{FD}_{j} ,
\label{eq2b26}
\end{eqnarray} 
in which 
\begin{eqnarray} 
{\cal F}_{j} = {\langle F\rangle_{j}\over F_{M}}, \quad
F_M = {\hbar\sqrt{\nu_{en}(\nu_{en}+\nu_{\rm imp})}\over el} ,  \quad
\tau_{e}= \sqrt{{\nu_{en}+\nu_{\rm imp}\over \nu_{en}}}. \label{eq2b27}
\end{eqnarray} 
Since $f^{FD}$ is an even function of $k$, its Fourier coefficient $f^{FD}_{j}$ is real. 
Note that equation~(\ref{3}) implies $f^{(0)}_{0} = f^{FD}_{0} = n$. 

The Chapman-Enskog ansatz consists of writing the distribution function as an expansion in
powers of the book-keeping parameter $\lambda$ (recall that we have to set $\lambda =1$
after retaining the desired number of terms) \cite{BEP03}
\begin{eqnarray} 
&& f(x,k,t; \lambda) = f^{(0)}(k;F) + \sum_{m=1}^{\infty} f^{(m)}(k;F)\, 
\lambda^{m} ,    \label{eq2b28}\\
&&  \varepsilon {\partial F\over\partial t} + \sum_{m=0}^{\infty}  J^{(m)}(F)\,
 \lambda^{m} = J(t).   \label{eq2b29}
\end{eqnarray} 
The coefficients $f^{(m)}(k;F)$ depend on the `slow variables' $x$ and $t$ only through 
their dependence on the electric field and the electron density (which are related through 
the Poisson equation). The field obeys a reduced evolution equation (\ref{eq2b29}), in 
which the functionals $J^{(m)}(F)$ are chosen so that the $f^{(m)}(k;F)$ are bounded and 
$2\pi/l$-periodic in $k$. Differentiating Amp\`ere's law (\ref{eq2b29}) with respect to 
$x$, we obtain the charge continuity equation. Moreover the condition
\begin{eqnarray} 
\int_{-\pi/l}^{\pi/l} f^{(m)}(k;n) \, dk = 2\pi\, f^{(m)}_{0}/l= 0,  \quad m\geq 1, 
\label{eq2b30}  
\end{eqnarray} 
ensures that $f^{(m)}$ for $m\geq 1$ does not contain contributions proportional to the 
zero-order term $f^{(0)}$. Note that the insertion of equation (\ref{eq2b28}) in Amp\`ere's 
law (\ref{eq2b25}) yields
\begin{eqnarray} 
J^{(m)} = {2e\over\hbar}\,\sum_{j= 1}^{\infty} j\langle \mbox{Im} [E_{1}(-j) 
f^{(m)}_{j}]\rangle_{j},     \label{eq2b31}
\end{eqnarray} 
which is also obtained by means of the above mentioned boundedness condition. 

Inserting equations~(\ref{eq2b28}) and (\ref{eq2b29}) into equation~(\ref{eq2b23}) and equating
all coefficients of $\lambda^m$ in the resulting series to zero, we find the hierarchy
\begin{eqnarray} 
{\cal L} f^{(1)} &=&  \left.  -
 \left({\partial f^{(0)}\over \partial t} + \sum_{j= -\infty}^{\infty} {i jl\over
 \hbar}\, e^{ijkl} E_{1}(j)\, {\partial\over \partial x}\langle f^{(0)}\rangle_{j}
\right) \right|_{0} , \label{eq2b32}\\
 {\cal L} f^{(2)} &=&  \left.  \left. -
 \left({\partial f^{(1)}\over \partial t} + \sum_{j= -\infty}^{\infty} {i jl\over
 \hbar}\, e^{ijkl} E_{1}(j)\, {\partial\over \partial x}\langle f^{(1)}\rangle_{j}
\right) \right|_{0} - {\partial \over \partial t} f^{(0)}\right|_{1}, \quad 
\label{eq2b33}
\end{eqnarray} 
and so on. We have defined 
\begin{eqnarray} 
{\cal L} u(k) \equiv {ie\over\hbar}\, \sum_{j=-\infty}^\infty jl\langle F
\rangle_{j} u_{j} e^{ijkl} + \left( \nu_{en} + {\nu_{\rm imp}\over 2}\right) u(k) 
- {\nu_{\rm imp} u(-k)\over 2}, \label{eq2b34}
\end{eqnarray} 
and the subscripts 0 and 1 in the right hand side of these equations mean that $\varepsilon\,
\partial F/\partial t$  is replaced by $J - J^{(0)}(F)$ and by $-J^{(1)}(F)$, respectively. 

 The linear equation ${\cal L} u= S$  has a bounded $2\pi/l$-periodic solution provided 
 $\int_{-\pi/l}^{\pi/l}S\, dk =0$. Equation~(\ref{eq2b32}) and this solvability 
condition yield equation~(\ref{eq2b31}) for $m=0$. The solution of equation~(\ref{eq2b32}) is 
\begin{eqnarray} 
f^{(1)}(k;F) = \sum_{j=-\infty}^{\infty} {\mbox{Re}S^{(1)}_{j} + i\,\tau_{e}^{-2}
\mbox{Im}S^{(1)}_{j} - ij{\cal F}_{j}S^{(1)}_{j}/\tau_{e}\over (1 + j^2 {\cal F}_{j
}^{2})\, \nu_{en}}\,  e^{ijkl},      \label{eq2b35}
\end{eqnarray} 
in which $S^{(1)}_{j}$ is the $j$th Fourier coefficient of the right hand side of 
equation~(\ref{eq2b32}). Using equation (\ref{eq2b35}), we can now explicitly write two terms in 
equation~(\ref{eq2b29}), thereby obtaining the following QDDE for the field and the electron
density given by the Poisson equation (\ref{2}) (with $\partial^2 W/\partial x^2
= \partial F/\partial x$). For the tight-binding dispersion relation, the QDDE is
\begin{eqnarray}
&&\varepsilon {\partial F\over\partial t} + {eN_{D}\over l}\,
{\cal N}\left(F,{\partial F\over\partial x}\right)\nonumber\\  
&& \quad = \varepsilon\,  \left\langle D\left(F,{\partial F\over\partial x},
{\partial^2 F\over \partial x^2}\right)\right\rangle_{1}
 + \left\langle A\left(F,{\partial F\over\partial x}\right)\right\rangle_{1}
 J(t) , \quad \quad \label{eq2b36}\\
&& A = 1 + {2 e v_{M}\over \varepsilon F_{M} l (\nu_{en}+ \nu_{\rm imp})}\, 
 {1- (1+2 \tau^{2}_{e})\, {\cal F}^2\over (1+ {\cal F}^{2})^3 }\, n {\cal M},  
\label{eq2b37}\\
&& {\cal N} = \langle n V {\cal M}\rangle_{1} + \langle (A - 1) 
\langle\langle n V {\cal M}\rangle_{1}\rangle_{1}\rangle_{1}\nonumber\\
&& \quad - { l \tau_{e}\Delta_1\over  F_{M}\hbar (\nu_{en}+
\nu_{\rm imp})}\,\left\langle {B\over 1+{\cal F}^2}\right
\rangle_{1} ,    \label{eq2b38}\\
&& V({\cal F}) = {2{\cal F}\over 1 + {\cal F}^2} , \quad
v_{M} = { l\, {\cal I}_{1}(M)\Delta_1\over 4\hbar\tau_{e} {\cal I}_{0}(M)},  
\label{eq2b39}\\
&& D = {l^2 \Delta_1^2 \over 8\hbar^2 (\nu_{en}+\nu_{\rm imp}) (1 + {\cal F}^{2}) }  
 \left( {\partial^2\langle F\rangle_{1}\over\partial x^2} - {4\hbar v_{M}
 \tau_{e} C\over l\Delta_1} \right) ,   \label{eq2b40}\\
&& B =  \left\langle {4 {\cal F}_{2}n{\cal M}_{2} 
\over (1+4{\cal F}^2_{2})^2} {\partial \langle F\rangle_{2}\over\partial x}
\right\rangle_{1} \nonumber\\
&& \quad + {\cal F} \left\langle {n{\cal M}_{2} (1-4{\cal F}^2_{2})
\over (1+4{\cal F}^2_{2})^2} {\partial \langle F\rangle_{2}\over\partial x}
\right\rangle_{1} \nonumber\\
&&  - {4\hbar v_{M} (1+\tau_{e}^2){\cal F}(n{\cal M})'\over l \tau_{e}
 (1+{\cal F}^2)\,\Delta_1}\left\langle n{\cal M} {1-{\cal F}^{2}\over 
 (1+ {\cal F}^2)^2} {\partial \langle F\rangle_{1}\over \partial x} 
 \right\rangle_{1}  \label{eq2b41}\\
&& C = \left\langle { (n {\cal M}_{2})' \over 1 +  4 {\cal F}_{2}^2 }\,
{\partial^2 F\over \partial x^2}\right\rangle_{1} - 2 {\cal F}\left\langle
{ (n {\cal M}_{2})' {\cal F}_{2}\over 1 +  4 {\cal F}_{2}^2 }\,
{\partial^2 F\over \partial x^2}\right\rangle_{1} \nonumber\\
&& \quad + {8\hbar v_{M} (1+\tau_{e}^2)
(n {\cal M})'\, {\cal F}\over l \tau_{e}\, (1 + {\cal F}^{2})\,\Delta_1}
\,\left\langle  {(n {\cal M})'{\cal F}\over 1 + {\cal F}^{2}}\,
{\partial^2 F\over \partial x^2} \right\rangle_{1}.   \label{eq2b42}\\
&& {\cal M}(n/N_{D}) = {{\cal I}_{1}(\tilde{\mu})\, {\cal I}_{0}(M)\over 
{\cal I}_{0}(\tilde{\mu}) \, {\cal I}_{1}(M)},  \quad 
{\cal M}_{2}(n/N_{D})= {{\cal I}_{2}(\tilde{\mu})\, {\cal I}_{0}(M)\over 
{\cal I}_{0}(\tilde{\mu}) \, {\cal I}_{1}(M)},\label{eq2b43}\\
&& {\cal I}_{m}(\tilde{\mu}) = {1\over \pi}\int_{-\pi}^{\pi}\cos (m k)
\left[\tan^{-1}\left(\frac{\tilde{\Gamma}/\delta}{1-\cos k}\right)\right.
\nonumber\\
&& \quad \quad \quad\left. +\int_{0}^\infty 
\frac{\tilde{\Gamma}}{(\tilde{E}-\delta+\delta\cos k)^2
+\tilde{\Gamma}^2}\,\ln\left( 1+e^{\tilde{\mu}-\tilde{E}}\right) d\tilde{E}
\right]\,dk.    \label{eq2b44}
\end{eqnarray} 
Here $g'$ denotes $dg/dn$, $\delta = \Delta_1/(2k_{B}T)$, $\tilde{\mu}=\mu/(k_{B}T)$, 
$\tilde{\Gamma}=\Gamma/(k_{B}T)$, ${\cal F}={\cal F}_{1}$, and $n=N_{D}$
at the particular value of the dimensionless chemical potential $\tilde{\mu}=M$. If the 
electric field and the electron 
density do not change appreciably over two SL periods, $\langle F\rangle_{j}\approx F$, the 
spatial averages can be ignored, and the {\em non-local}  QDDE (\ref{eq2b36}) becomes the 
{\em local} generalized DDE (GDDE) obtained from the semiclassical theory 
\cite{BEP03}. The boundary conditions for the QDDE (\ref{eq2b36}), which contains 
triple spatial averages, need to be specified for the intervals $[-2l,0]$ and $[Nl,Nl+2l]$, 
and not just at the points $x=0$ and $x=Nl$ ($N$ denotes the number of SL periods spanning
the device), as in the case of the parabolic semiclassical GDDE. Similarly, the initial 
condition has to be defined on the extended interval $[-2l,Nl+2l]$. Note that the spatial
averages in the nonlocal QDDE give rise to finite differences of partial derivatives in the
diffusion terms, and therefore lead to a type of equations for which little seems to be known.

\vspace*{1pt}\textlineskip 
\setcounter{equation}{0}
\section{Numerical solution of the QDDE} 
\vspace*{-0.5pt}

\noindent In this Section, we solve the nonlocal QDDE (\ref{eq2b36}) together with
the voltage bias condition
\begin{equation}
\int_{0}^{Nl} F(x,t)\, dx = \phi F_{M} Nl,   \label{n1}
\end{equation}
for the field $F$ and the total current density $J$. As boundary conditions in the intervals
$[-2l,0]$ and $[Nl,Nl+2l]$, we adopt
\begin{equation}
J-\varepsilon\,\frac{\partial F}{\partial t}= \sigma F,   \label{n2}
\end{equation}
at all points $[-2l,0]$ of the Ohmic injecting contact and zero-flux boundary conditions at
the receiving contact $[Nl,Nl+2l]$. The contact conductivity $\sigma$ is selected so that
$e N_{D}v_{M} V(F/F_{M})$ and $\sigma F$ intersect on the second branch of $V({\cal
F})$, in which $dV/d{\cal F}<0$. This is a typical boundary condition yielding
self-sustained oscillations in drift-diffusion SL models \cite{Bon02,BEP03}. We have
used a constant initial condition $F=\phi F_{M}$ in our numerical simulations. The SL
parameters we have used correspond to a 157-period 3.64~nm GaAs/0.93~nm AlAs SL 
\cite{sch98} at 14K, with $\Delta=72$ meV, $N_D = 4.57 \times 10^{10}$ cm$^{-2}$, 
$\nu_{\rm imp} = 2\nu_{en} = 18\times 10^{12}$ Hz under a dc voltage bias of 1.62 V
($\phi=1$). Cathode and anode contact conductivities are 2.5 and 0.62 $\Omega^{-1}
\mbox{cm}^{-1}$, respectively, and the effective mass is $m^*= (0.067 d_{W} + 0.15 
d_{B}) m_0/l$, where $m_0=9.109534 \times 10^{-31}$Kg is the electron rest mass.

\begin{figure}[ht]
\begin{center}
       \epsfxsize=120mm
       \epsfbox{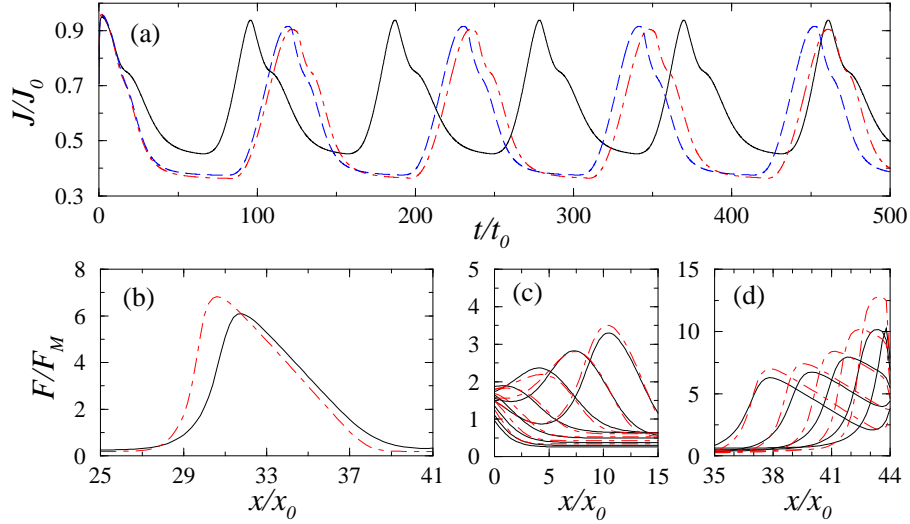}
\caption{
(a) Current ($J_{0}=ev_{M}N_{D}/l$) vs.\ time ($t_{0}=\varepsilon F_{M}/J_{0}$) 
during self-oscillations for a voltage biased GaAs/AlAs SL, as described by the 
QDDE (solid line), the QDDE with $\Gamma=0$ (long-dashed line) and by the GDDE 
(dot-dashed line).  (b) Comparison between the fully developed dipole wave for the QDDE
(solid line) and the dipole wave for the GDDE (dashed line). (c) Dipole wave at different
times during the stage in which it is shed from the injecting contact. (d) Same as (c) for the
stage in which the dipole disappears at the anode, located at $Nl/x_0 \approx 44$. Parameter 
values are $x_{0}=v_{M}t_{0}=16$ nm, $t_{0}= 0.43$ ps, $J_{0}= 6.07 \times 10^5$ 
A/cm$^2$, $\phi=1$.}
\label{fig}
\end{center}
\end{figure}

Our numerical solution shows that the current and the field profile become stationary for 
$\phi< 0.75$ (1.2 V). For larger values of the dimensionless voltage $\phi$, the initial field 
profile evolves toward a stable time-periodic solution for which $J$ oscillates with time and 
the field profile shows recycling and motion of a pulse from $x=0$ to the SL end. 
Fig.~\ref{fig} shows the self-oscillations of the current for $1.62$ V ($\phi=1$) and the 
corresponding field pulse at different times. In this figure, we compare the solution of the 
GDDE corresponding to the semiclassical BGK-Poisson kinetic equation and the solution 
of the QDDE for $\Gamma=0$ (no collision broadening) and for $\Gamma=18$ meV which is 
of the same order as the collision frequencies.

Self-oscillations in the QDDE have a frequency of $\nu_Q = 25.5$ Ghz, faster than in the GDDE,
$\nu_G = 20.6$ Ghz (relative frequency $(\nu_Q - \nu_G)/\nu_G = 23.8 \%$). When $\Gamma=0$, the
frequency is $\nu_{Q^*} = 21$ Ghz, and the relative frequency $(\nu_{Q^*} - \nu_G)/\nu_G = 1.94 \%$.

Collision broadening shortens the period of the current oscillations and therefore it reinforces 
the effects of the nonlocal terms in the QDDE due to quantum effects.

\vspace*{1pt}\textlineskip 
\setcounter{equation}{0}
\section{Discussion} 
\vspace*{-0.5pt}

\noindent We have derived a nonlocal drift-diffusion model for a strongly coupled
superlattice with one miniband by applying the Chapman-Enskog procedure to a simple
WPBGK quantum kinetic equation. Our local equilibrium function is a collision-broadened
Fermi-Dirac distribution inspired in the well-known thermal equilibrium of the
Kadanoff-Baym equations \cite{KBabook89,Wac02}, but the effects of the electric field
on the collision operator have been ignored. For the case of a quantum particle 
in an external potential, Degond and Ringhofer \cite{DRi03} have proposed a different 
form of the BGK collision model, in which the density matrix corresponding to local
 equilibrium is a Maxwellian operator obtained by minimizing the Boltzmann entropy subject 
 to moment constraints. Their formulation includes the effects of the electric field on the 
collisions, and the chemical potential is a nonlocal functional of the electron density, not a
function given by solving Eq.\ (\ref{3}) for $\mu$. This 
functional relation introduces additional nonlocalities in the balance equations resulting from 
the Chapman-Enskog procedure. See \cite{DMR04}, in which quantum drift-diffusion and
energy-transport models are derived for the case of a particle in an external potential
undergoing collisions according to their BGK model. Notice that the BGK collision models
used by these authors conserve kinetic energy, which is a somewhat unrealistic feature for 
semiconductors. 

In this paper, we have treated the case of a one-dimensional SL with one miniband. We have
used a Wigner-Poisson equation with relaxation towards a broadened local Fermi-Dirac
distribution and a simple elastic impurity collision term. We have derived a nonlocal 
drift-diffusion equation and solved it numerically to show that our model displays the
self-oscillations of the current which are observed in experiments. There are several
directions in which we could extend our results. Firstly, it would be interesting to derive
quantum hydrodynamic models by using a more general BGK collision model and compare
it with the models by Lei and collaborators \cite{LTi84,Lei95,CLe99}. Our approximation
$f^{(0)}$ is possibly closer to the real electron distribution than the parametrized local
equilibrium used by Lei et al, which has been criticized by other authors \cite{Wac02}.
Unlike Lei's, our approximation $f^{(0)}$ takes explicitly into account the effect of the 
electric field (and therefore the modified ``equilibrium'' $f^{(0)}$ corrects somewhat the 
effect of having neglected the electric potential in our collision model). Secondly, it would be 
important to consider the case of a SL with several minibands, although figuring out a 
reasonable local equilibrium distribution to include in
the BGK collision model is probably harder. Thirdly, there are many interesting 
mathematical questions arising from these derivations. To mention one, the trailing edge of 
the pulse in Fig.~\ref{fig} is rather sharp. In related work with drift-diffusion models of the
Gunn effect, it is known that this trailing edge is close to a shock wave of the zero-order 
hyperbolic drift equation \cite{Bon91}. Obviously, the diffusive terms in the QDDE
regularize this shock, and so do the collision terms in the kinetic equation (even if we think 
in terms of the semiclassical kinetic equation). What is the relation between these two 
regularizations? Caflisch has considered a related problem in gas dynamics \cite{Caf83}. 
We hope that our work paves the way to tackling these problems in the future.


\nonumsection{Acknowledgment}
\noindent
This work has been 
supported by the MCyT grant BFM2002-04127-C02-01, and by the 
European Union under grant HPRN-CT-2002-00282. R.E. has been supported by a 
postdoctoral grant awarded by the Consejer\'\i a de Educaci\'on of the Autonomous Region 
of Madrid.



\nonumsection{References}

\end{document}